\documentclass[conference]{IEEEtran}
\IEEEoverridecommandlockouts
\usepackage{amsfonts}
\usepackage{dsfont}
\usepackage{amssymb}
\usepackage{graphicx}
\usepackage{subfigure}
\usepackage{enumerate}
\usepackage{amsmath}
\usepackage{color}
\usepackage{amsthm}
\usepackage{amsmath}
\usepackage{algorithm}
\usepackage{algpseudocode}
\usepackage[bookmarks=false]{hyperref}
\hypersetup{hidelinks}
\usepackage{breakurl}
\usepackage{cite}
\usepackage{epstopdf}

\usepackage{verbatim}

\allowdisplaybreaks[4]

\newcommand{\bp}{\begin{proof} \small }
\newcommand{\ep}{\end{proof} \normalsize}
\newcommand{\epx}{\end{proof} \small}
\newcommand{\bpa}{\begin{proofappx} \footnotesize }
\newcommand{\epa}{\end{proofappx} \small }
\newtheorem{theorem}{Theorem}

\newtheorem{lemma}{Lemma}

\newtheorem*{theorem*}{Theorem}
\newtheorem*{proposition*}{Proposition}
\newtheorem*{corollary*}{Corollary}
\newtheorem*{lemma*}{Lemma}
\newtheorem*{assumption*}{Assumption}
\newtheorem*{definition*}{Definition}
\newtheorem*{claim*}{Claim}

\newcommand{\be}{\begin{equation}}
\newcommand{\ee}{\end{equation}}
\newcommand{\bs}{\begin{subequations}}
\newcommand{\es}{\end{subequations}}
\newcommand{\bq}{\begin{eqnarray}}
\newcommand{\eq}{\end{eqnarray}}
\newcommand{\bqn}{\begin{eqnarray*}}
\newcommand{\eqn}{\end{eqnarray*}}

\newcommand{\ba}{\left[ \begin{array}}
\newcommand{\ea}{\\ \end{array} \right]}
\newcommand{\ben}{\begin{enumerate}}
\newcommand{\een}{\end{enumerate}}

\def\e{{\boldsymbol{e}}}

\def\g{{\boldsymbol{g}}}

\def\m{{\boldsymbol{m}}}

\def\v{{\boldsymbol{v}}}
\def\w{{\boldsymbol{w}}}
\def\x{{\boldsymbol{x}}}
\def\y{{\boldsymbol{y}}}
\def\z{{\boldsymbol{z}}}

\def\real{{\mathchoice%
{\hbox{\rm\setbox1=\hbox{I}\copy1\kern-.45\wd1 R}}
{\hbox{\rm\setbox1=\hbox{I}\copy1\kern-.45\wd1 R}}
{\hbox{\scriptsize\rm\setbox1=\hbox{I}\copy1\kern-.45\wd1 R}}
{\hbox{\scriptsize\rm\setbox1=\hbox{I}\copy1\kern-.45\wd1 R}}}}

\def\Zint{{\mathchoice{\setbox1=\hbox{\sf Z}\copy1\kern-.75\wd1\box1}
{\setbox1=\hbox{\sf Z}\copy1\kern-.75\wd1\box1}
{\setbox1=\hbox{\scriptsize\sf Z}\copy1\kern-.75\wd1\box1}
{\setbox1=\hbox{\scriptsize\sf Z}\copy1\kern-.75\wd1\box1}}}
\newcommand{\complex}{ \hbox{\rm C\kern-0.45em\rule[.07em]{.02em}{.58em}%
\kern 0.43em}}

\begin{document}

\title{Time-Correlated Sparsification for \\Efficient Over-the-Air Model Aggregation in \\Wireless Federated Learning}

\author{\IEEEauthorblockN{Yuxuan Sun$^*$, Sheng Zhou$^*$, Zhisheng Niu$^*$, Deniz G\"und\"uz$^\dagger$}
	\IEEEauthorblockA{$^*$Beijing National Research Center for Information Science and Technology\\
		Department of Electronic Engineering, Tsinghua University, Beijing 100084, China\\
		$^\dagger$Department of Electrical and Electronic Engineering, Imperial College London, London SW7 2BT, UK\\
		Email: \{sunyuxuan, sheng.zhou, niuzhs\}@tsinghua.edu.cn, d.gunduz@imperial.ac.uk}
}
	
\maketitle

\begin{abstract}
	Federated edge learning (FEEL) is a promising distributed machine learning (ML) framework to drive edge intelligence applications. However, due to the dynamic wireless environments and the resource limitations of edge devices, communication becomes a major bottleneck.
	In this work, we propose time-correlated sparsification with hybrid aggregation (TCS-H) for communication-efficient FEEL, which exploits jointly the power of model compression and over-the-air computation. By exploiting the temporal correlations among model parameters, we construct a global sparsification mask, which is identical across devices, and thus enables efficient model aggregation over-the-air. Each device further constructs a local sparse vector to explore its own important parameters, which are aggregated via digital communication with orthogonal multiple access. We further design device scheduling and power allocation algorithms for TCS-H. Experiment results show that, under limited communication resources, TCS-H can achieve significantly higher accuracy compared to the conventional top-K sparsification with orthogonal model aggregation, with both i.i.d. and non-i.i.d. data distributions.

\end{abstract}

%
	
%
\IEEEpeerreviewmaketitle

\section{Introduction}

Federated edge learning (FEEL) refers to the implementation of federated learning algorithms \cite{googleai,McMahan2017commun} in a wireless network, where edge devices train a shared machine learning (ML) model using their local datasets, and periodically communicate with a base station (BS) via wireless channels for global model aggregation.
It is considered as a promising paradigm to facilitate edge intelligence, driving numerous applications such as Internet of things, augmented and virtual reality, self driving, and smart network management \cite{Park2019wireless,Chen2021distributed}. 

However, due to resource limitations in terms of wireless channel bandwidth, transmit power, battery capacity, and computing capability of edge devices, it is challenging to train accurate models in a FEEL system.
A particular challenge is the communication bottleneck, which limits the amount of information that can be exchanged between the BS and edge devices, and thus reduces the training performance. 

To overcome the communication bottleneck, a straightforward idea is to reduce the traffic load. 
Model compression techniques, mainly quantization and sparsification, are proposed to reduce the communication load \cite{QSGD,signSGD,Stich2018Sparsified,Basu2019Qsparse,Emre2021spars}. With quantization \cite{QSGD}, fewer bits, or even a single bit \cite{signSGD}, are used to represent each parameter in the ML model.
With sparsification, the high dimensional ML model vector is transformed to a sparse vector, where only important elements are kept \cite{Stich2018Sparsified}. The two techniques can also be combined \cite{Basu2019Qsparse}. While a high dimensional mask is required to be transmitted from each device to the BS for conventional sparsification methods, it is shown in \cite{Emre2021spars} that a global mask can be used instead by exploiting the temporal correlations during the convergence of the ML model.
Communication load can also be reduced by enabling multiple iterations of local training, and carefully scheduling a subset of devices in each round \cite{yang2019scheduling, Shi2021Joint}. 

Another emerging way of improving the communication efficiency of FEEL is over-the-air computation (OAC) \cite{zhu2019broadband,mma2019federated, mma2020machine, Sun2020edge, Sun2021JSAC}, which exploits the superposition property of the wireless multiple access channel from devices to the BS to average the local models over-the-air. It is shown in \cite{zhu2019broadband} that compared with the digital counterpart, an over-the-air FEEL system with $N$ devices can improve the communication efficiency by $O\left(\frac{N}{\log_2 N}\right)$. However, due to the high dimension of the ML model, it is still not practical for the devices to update all the parameters to the BS at each time. A linear projection method is proposed in \cite{mma2019federated, mma2020machine}, which projects the model vector to a lower-dimension by a pseudo-random matrix. However, this increases the computational complexity at the receiver, making it difficult for real implementation.
Temporal correlations of model parameters are further exploited in \cite{Fan2021Temporal} to reduce the recovery complexity.

In this paper, we combine the power of model compression with OAC, and propose time-correlated sparsification with hybrid aggregation (TCS-H) to further improve the communication efficiency of FEEL. 
Our algorithm is built upon the time-correlated sparsification (TCS) method in \cite{Emre2021spars}, which exploits the temporal correlations in the model parameters during training, and constructs a global mask for sparsification.
As the global mask is identical across devices, we propose to aggregate the corresponding sparsified model differences over-the-air. Meanwhile, local sparsification is used to explore the important parameters specific to each device. With quantization, these sparse vectors, together with the local sparsification masks, are transmitted via digital communication with orthogonal multiple access. Under an average power constraint, we further design the device scheduling and power allocation algorithm. 
Experiment results on CIFAR-10 dataset show that, under limited communication resources, TCS-H can improve the accuracy by $6.6\%$, compared with the classical Top-K sparsification with orthogonal model aggregation.

\section{Time-Correlated Sparsification with Hybrid Aggregation for FEEL}

\begin{figure}[!t]
	\centering
	\includegraphics[width=0.48\textwidth]{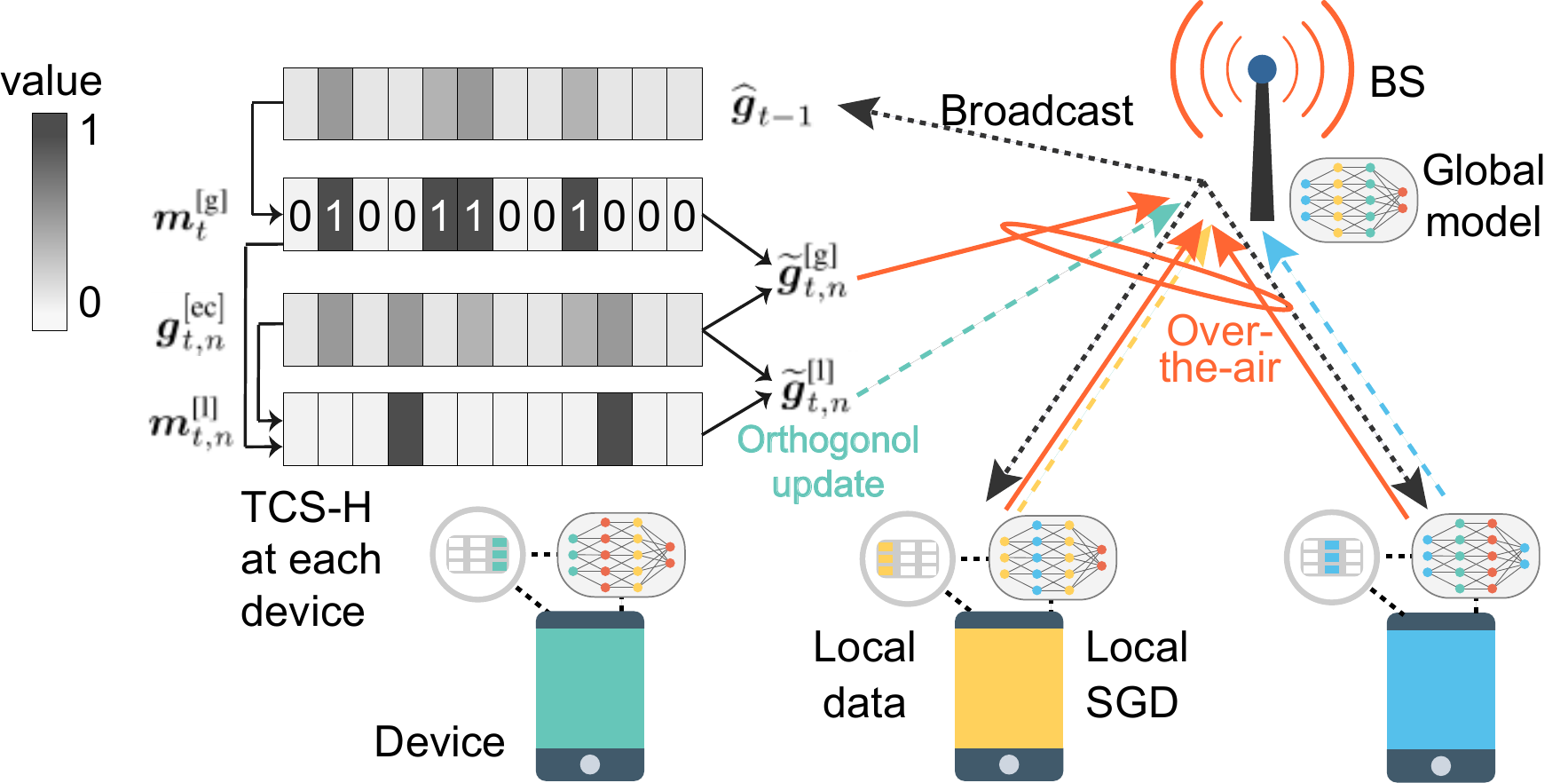}
	\vspace{-2mm}	
	\caption{Illustration of the FEEL system and the proposed TCS-H algorithm.}
	\label{system}
	\vspace{-6mm}
\end{figure}

As shown in Fig. \ref{system}, we consider a FEEL system where $N$ devices $\mathcal{N}=\{1,\ldots,N\}$ train a ML model collaboratively, under the coordination of a BS. Each device $n$ owns a local dataset $\mathcal{D}_n$ with $D$ data samples. The local loss function is defined as $F_n(\w)\triangleq \frac{1}{D}\sum_{\boldsymbol{\xi}_i \in\mathcal{D}_n} f(\w,\boldsymbol{\xi}_i)$,
where $\boldsymbol{\xi}_i$ is a data sample in $\mathcal{D}_n$, $\w\in \mathbb{R}^d$ is the model vector to be trained, and $f(\w,\boldsymbol{\xi}_i)$ is a loss function measuring the fitting performance of $\w$ on a single data sample $\boldsymbol{\xi}_i$.
The global loss function is defined as $F(\w) \triangleq  \frac{1}{N}\sum_{n=1}^{N}F_n(\w)$. The goal of FEEL is $\min_{\w} F(\w)$, by optimizing the model vector $\w$.

To train the FEEL task, devices carry out stochastic gradient descent (SGD) locally, and communicate with the BS for global model aggregation in an iterative manner. 
There are $M$ orthogonal sub-channels for uplink communications, indexed by $\mathcal{M}=\{1,\ldots,M\}$. Each device $n$ has an average power constraint $\bar{P}_n$. A total number of $C$ time slots are available for uplink communications during the whole training process, where the length of a time slot equals that of a symbol. 
We remark here that, different from the common assumption that limits the training rounds, we mainly focus on the uplink communication bottleneck by limiting the total number of resource blocks to $MC$.
Accordingly, the total number of training rounds, denoted by $T_{\text{Algo}}$, depends on the FEEL algorithm and its parameter settings.

In order to optimize the training performance under the constraints on communication bandwidth and average transmit power, we aim to reduce the communication load while improving the efficiency of model aggregation. Combining model compression with OAC, we propose the TCS-H algorithm for FEEL, as shown in Algorithm \ref{AlgoTCS}.

\begin{algorithm}[t]
	\caption{TCS-H for FEEL}\label{AlgoTCS}
	\begin{algorithmic}[1]
		\State\textbf{Initialization:} Initial model $\w_{-1}$, initial global gradient $\widehat{\g}_{0}=\g_0=\w_0-\w_{-1}$. Let $U_0=0$, $t=1$.
		\While{$\sum_{\tau=0}^{t-1}U_{\tau}\leq C$ and model is not converged }
		\State BS broadcasts the global model difference $\widehat{\g}_{t-1}$ of last round to all the devices.
		\State \textbf{Local training at each device $n\in\mathcal{N}$: }
		\State Recovers the global model $\w_{t-1}\leftarrow\w_{t-2}+\widehat{\g}_{t-1}$.
		\State Carries out local SGD for $H$ iterations, and gets local model difference: $\g_{t,n}=\w_{t,n,H}-\w_{t-1}$.
		\State Error compensation: $\g_{t,n}^{[\text{ec}]}=\g_{t,n}+\e_{t-1,n}$.
		\State Global mask: $\m_t^{\text{[g]}}=S_{\text{top}}(\widehat{\g}_{t-1}, K^{\text{[g]}}) $.
		\State Sparsification with global mask: $\widetilde{\g}_{t,n}^{\text{[g]}}=\m_t^{\text{[g]}} \circ {\g}_{t,n}^{[\text{ec}]}$.
		\State Local mask: $\m_{t,n}^{\text{[l]}}=S_{\text{top}}\left( (\boldsymbol{1}-\m_t^{\text{[g]}})\circ{\g}_{t,n}^{[\text{ec}]}, K^{\text{[l]}}\right) $.
		\State Sparsification and quantization with local mask: $\widetilde{\g}_{t,n}^{\text{[l]}}= Q(\m_{t,n}^{\text{[l]}} \circ {\g}_{t,n}^{[\text{ec}]},q)$.
		\State \textbf{Model aggregation at the BS:}
		\State Device scheduling and power allocation according to Algorithm \ref{AlgoSchedule}. Records communication slots $U_t$.
		\State Aggregates $\widetilde{\g}_{t,n}^{\text{[g]}}$ of all the scheduled devices over-the-air, and receives $\widetilde{\g}_{t,n}^{\text{[l]}}$, $\m_{t,n}^{\text{[l]}}$ via orthogonal transmissions.
		\State Aggregates the global model difference $\widehat{\g}_t $ according to \eqref{gt_global}, and updates global model $\w_t \leftarrow \w_{t-1}+\widehat{\g}_t $.
		\State $t\leftarrow t+1$.
		\State \textbf{Error cumulation at each device $n\in\mathcal{N}$: }
		\If {$n\in\mathcal{N}_t$}
		\State $\e_{t,n}\leftarrow \g_{t,n}^{[\text{ec}]}-\widetilde{\g}_{t,n}^{\text{[g]}}-\widetilde{\g}_{t,n}^{\text{[l]}}$.
		\Else
		\State $\e_{t,n}\leftarrow \e_{t-1,n}$.
		\EndIf
		\EndWhile
	\end{algorithmic}
\end{algorithm}

In the initialization phase, the BS and all the devices start from a common model $\w_{-1}$. Each device calculates the local gradient using its own dataset, and the BS aggregates these to get the global gradient $\g_0$. As the initialization only involves a single round, the communication cost of the initial training phase is not taken into account.

In the main loop, there are three steps in each round.

\emph{1) Local SGD:}
In the $t$-th training round, each device receives the global model difference $\widehat{\g}_{t-1}$ from the BS, and updates the global model from the last round according to $\w_{t-1}=\w_{t-2}+\widehat{\g}_{t-1}$. Each device $n$ then carries out $H$ iterations of local SGD according to
\begin{align}
\w_{t,n,i} = \w_{t,n,i-1} - \eta_t \nabla F_n(\w_{t,n,i-1};\mathcal{B}_{t,n,i}), i=1,\ldots,H, \nonumber
\end{align}
where $\w_{t,n,0}=\w_{t-1}$, $\eta_t$ is the learning rate, and $\mathcal{B}_{t,n,i}$ is a randomly sampled mini-batch with batch size $B$.
The local model difference is then calculated according to $\g_{t,n}= \w_{t,n,H} -\w_{t-1}$, and compensated with the cumulated error $\e_{t-1,n}$, which yields $\g_{t,n}^{[\text{ec}]}=\g_{t,n}+\e_{t-1,n}$. The cumulated error keeps track of the parameters that are not updated to the BS due to compression, helping to accelerate the training \cite{Stich2018Sparsified}.

\emph{2) Local Model Compression:}
We use the TCS method \cite{Emre2021spars} as well as stochastic quantization to reduce the communication load in each round.
Let function $S_{\text{top}}(\x, K)$ return a mask vector $\m\in\{0,1\}^d$, which indicates the locations of non-zero elements after a top-$K$ sparsification operation on $\x$. That is, if the $i$-th element of $\x$ after a top-$K$ sparsification is not zero, then $\m[i]=1$; otherwise $\m[i]=0$. The top-$K$ sparsification keeps the $K$ elements of $\x$ with largest absolute values, while setting the remaining elements to zero.

Based on the global model difference of the last round, $\widehat{\g}_{t-1}$, each device first calculates the global mask $\m_t^{\text{[g]}}=S_{\text{top}}(\widehat{\g}_{t-1}, K^{\text{[g]}}) $ with sparsity $K^{\text{[g]}}$. Then, the local model difference ${\g}_{t,n}^{[\text{ec}]}$ is sparsified with the global mask, which yields $\widetilde{\g}_{t,n}^{\text{[g]}}=\m_t^{\text{[g]}} \circ {\g}_{t,n}^{[\text{ec}]}$, where $\circ$ represents the element-wise multiplication. By exploiting the temporal correlations of the model, the global mask is the same for all the devices. This enables OAC for global model aggregation.

Then, each device generates a local mask $\m_{t,n}$ to explore the important elements of ${\g}_{t,n}^{[\text{ec}]}$, where $\m_{t,n}^{\text{[l]}}=S_{\text{top}}\left( (\boldsymbol{1}-\m_t^{\text{[g]}})\circ{\g}_{t,n}^{[\text{ec}]}, K^{\text{[l]}}\right)$, and $\boldsymbol{1}$ is a $d$-dimension all-one vector. The local model difference is then sparsified with the local mask and then quantized, which is given by $\widetilde{\g}_{t,n}^{\text{[l]}}=  Q(\m_{t,n} ^{\text{[l]}} \circ {\g}_{t,n}^{[\text{ec}]},q)$. Here, $Q(\x, q)$ represents a $q$-bit stochastic quantization function, with examples in \cite{Basu2019Qsparse}.

\emph{3) Global Model Aggregation:}
As the positions of non-zero elements of $\widetilde{\g}_{t,n}^{\text{[g]}}$ are the same for all the devices, we propose to use OAC for the global aggregation of $\widetilde{\g}_{t,n}^{\text{[g]}}$ to improve the communication efficiency. Meanwhile, $\widetilde{\g}_{t,n}^{\text{[l]}}$ and local masks $\m_{t,n}^{\text{[l]}}$ are aggregated via digital communication with orthogonal multiple access; and thus, quantization is introduced above to further reduce the communication load.

Specifically, in the $t$-th round, a number of $N_t$ devices in set $\mathcal{N}_t\subset\mathcal{N}$ are scheduled for global aggregation according to Algorithm \ref{AlgoSchedule}, which will be introduced in the next section.
For OAC, each device $n\in\mathcal{N}_t$ extracts the non-zero elements of $\widetilde{\g}_{t,n}^{\text{[g]}}$ and partitions them evenly into $M$ segments $\left[\widetilde{\g}_{t,n,1}^{\text{[g]}}, \cdots, \widetilde{\g}_{t,n,M}^{\text{[g]}} \right]$, each with $\left\lceil\frac{K^{\text{[g]}}}{M}\right\rceil$ or $\left \lfloor\frac{K^{\text{[g]}}}{M}\right\rfloor$ elements. 
The wireless channel gain between the BS and device $n$ in the $m$-th sub-channel is denoted by $h_{t,n,m}$, which is assumed to be constant during each round. 
In the $m$-th sub-channel, each scheduled device transmits $\frac{\sigma_t \widetilde{\g}_{t,n,m}^ {\text{[g]}}} {h_{t,n,m}}$ synchronously with all the other scheduled devices for OAC, where $\sigma_t$ is a power scalar that determines the received signal-to-noise ratio (SNR).


At the BS, the received signal over sub-channel $m$ is 
\begin{align}
\y_{t,m} \!\!=\!\!\sum_{n\in\mathcal{N}_t} \!\!h_{t,n,m} \! \frac{\sigma_t \widetilde{\g}_{t,n,m}^{\text{[g]}}}{h_{t,m,n}} \!+ \!\z_{t,m} 
\!=\! \sigma_t \!\!\sum_{n\in\mathcal{N}_t}\!\widetilde{\g}_{t,n,m}^{\text{[g]}} \!+\! \z_{t,m},
\end{align}
where $\z_{t,m}$ is a noise vector, with each element following Gaussian distribution with zero mean and variance $\sigma_0^2$. 
Let $j(i)$ be the index of the $i$-th non-zero element of $\m_t^{\text{[g]}}$. The BS constructs a $d$-dimension sparse vector $\y_t\in\mathbb{R}^d$ based on $\m_t^{\text{[g]}}$, where $\y_t[j(i)]$ equals the $i$-th element of $[\y_{t,1},\cdots, \y_{t,M}]$. Likewise, $[\z_{t,1},\cdots, \z_{t,M}]$ is mapped to $\z_t\in\mathbb{R}^d$.


In the digital communication part, each device $n\in\mathcal{N}$ transmits both $\m_{t,n}^{\text{[l]}}$  and the non-zero elements of $\widetilde{\g}_{t,n}^{\text{[l]}}$ to the BS, requiring $Q^{\text{[l]}}=(\log_2 d+q)K^{\text{[l]}}$ bits. The sub-channel assignment and the corresponding power allocation are given by Algorithm \ref{AlgoSchedule}.

The global model difference $\widehat{\g}_t $ is aggregated at the BS: 
\begin{align}\label{gt_global}
\widehat{\g}_t =& \frac{\y_t}{\sigma_tN_t}+\frac{1}{N_t}\sum_{n\in\mathcal{N}_t}\widetilde{\g}_{t,n}^{\text{[l]}} \nonumber\\
=&\frac{1}{N_t}\sum_{n\in\mathcal{N}_t}\widetilde{\g}_{t,n}^{\text{[g]}} +\frac{\z_t}{\sigma_t N_t} +\frac{1}{N_t}\sum_{n\in\mathcal{N}_t}\widetilde{\g}_{t,n}^{\text{[l]}} .
\end{align}

Finally, error cumulation is carried out at each device. If device $n$ is scheduled, the difference between the transmitted vector and the original vector  is accumulated, i.e., $\e_{t,n}= \g_{t,n}+\e_{t-1,n}-\widetilde{\g}_{t,n}^{\text{[g]}}-\widetilde{\g}_{t,n}^{\text{[l]}}$. Otherwise, $\e_{t,n}= \e_{t-1,n}$.
Define $U_t$ as the total number of communication time slots used in the $t$-th round. If the global model is not converged and the communication resource is not used up, i.e., $\sum_{\tau=1}^{t} U_{\tau} < C$, the TCS-H algorithm will start the next round.

\section{Device Scheduling and Power Allocation}

In this section, we introduce the device scheduling and power allocation algorithm for TCS-H.
We consider that each device has an average power constraint $\bar{P}_n$ for communication. For OAC, the total power required in the $t$-th round at device $n$ is $P_{t,n}^{\text{[g]}}=\sum_{m=1}^{M}\left\Vert\frac{\sigma_t\widetilde{\g}_{t,n,m}^{\text{[g]}}}{h_{t,n,m}}\right\Vert_2^2$,
with $U_t^{\text{[g]}}=\left\lceil\frac{K^{\text{[g]}}}{M}\right\rceil$ communication time slots. For digital communication, define $P_{t,n,m}^{\text{[l]}}$ as the power allocated to sub-channel $m$ by device $n$ in round $t$, and denote the corresponding required slots by $U_{t,n}^{\text{[l]}}$.
Then, the average power constraint is given by $\frac{1}{C}\sum_{t=1}^{T_{\text{Algo}}}\left(P_{t,n}^{\text{[g]}}+U_{t,n}^{\text{[l]}} \sum_{m=1}^{M}P_{t,n,m}^{\text{[l]}}\right)\leq \bar{P}_n$.

Let	$P_{t,n}^{\text{[l]}}= \sum_{m=1}^{M} P_{t,n,m}^{\text{[l]}}$, and $P_{t,n}=P_{t,n}^{\text{[g]}}+U_{t,n}^{\text{[l]}}P_{t,n}^{\text{[l]}}$. Then, at the start of round $t$, the average power constraint of the $n$-th device is given by $\bar{P}_{t,n}=\frac{C\bar{P}_n-\sum_{\tau=1}^{t-1}P_{\tau,n}}{C-\sum_{\tau=1}^{t-1}U_{\tau}}$.

\emph{Device Scheduling:}
As shown in Algorithm \ref{AlgoSchedule}, after local gradient computation in the $t$-th round, devices that satisfy the power constraint $P_{t,n}^{\text{[g]}}\leq \alpha_{t,n}\bar{P}_{t,n}U_{t,n}^{\text{[g]}}$ are scheduled by the BS for global aggregation. Here, $\alpha_{t,n}>0$ is a coefficient which dynamically adjusts the power allocation policy. $\alpha_{t,n}=1$ indicates a simple myopic policy. If $\alpha_{t,n}>1$, power is allocated in a more aggressive manner in the current round, and vice versa. 
In this work, we consider a myopic policy and do not further optimize this term, leaving it as a future work. We remark that, the selection of $\alpha_{t,n}$ requires an online device scheduling and power allocation algorithm, which can be designed based on stochastic optimization theories such as Lyapunov optimization. We refer to our previous work \cite{Sun2021JSAC} for such a design approach.


\emph{Sub-Channel Assignment and Power Allocation:}
For all the scheduled devices $n\in\mathcal{N}_t$, the BS needs to assign sub-channels and allocate power for them to aggregate $\widetilde{\g}_{t,n}^{\text{[l]}}$ and $\m_{t,n}^{\text{[l]}}$. We consider that each sub-channel is allocated to one device in each round. The goal is to minimize the required communication time slots $U_{t}^{\text{[l]}}\triangleq \max_{n\in\mathcal{N}_t}U_{t,n}^{\text{[l]}}$, under the power constraints of the scheduled devices. Let $\beta_{t,n,m}=1$ indicate that the $m$-th sub-channel is assigned to the $n$-th device for digital communication in round $t$, and $\beta_{t,n,m}=0$ otherwise. Let $\boldsymbol{\beta}_t=\{\beta_{t,n,m}| n\in\mathcal{N}_t, m\in\mathcal{M}\}$,
and $\boldsymbol{P}_t=\{P_{t,n,m}^{\text{[l]}}| n\in\mathcal{N}_t, m\in\mathcal{M}\}$. The problem is:
\begin{subequations}
	\begin{align}
	\mathcal{P}1:~\min_{ \boldsymbol{\beta}_t,\boldsymbol{P}_t} &~\max_{n\in\mathcal{N}_t}U_{t,n}^{\text{[l]}} \\
	\text{s.t.} ~&~U_{t,n}^{\text{[l]}}  R_{t,n} \geq Q^{\text{[l]}}, \forall n \in\mathcal{N}_t, \label{cons_bit}\\
	&\sum_{m=1}^{M} P_{t,n,m}^{\text{[l]}} \leq \bar{P}_{t,n}, \forall n \in\mathcal{N}_t, \label{cons_power}\\
	&\sum_{n\in\mathcal{N}_t}\beta_{t,n,m} \leq 1, \forall m \!\in\! \mathcal{M}, ~\beta_{t,n,m}\!\in\!\{0,1\},\label{cons_beta}
	\end{align}
\end{subequations}
where $R_{t,n}=\sum_{m=1}^{M} \beta_{t,n,m}\log_2\left( 1+\frac{P_{t,n,m}^{\text{[l]}}|h_{t,n,m}|^2}{\sigma_0^2}  \right)$ is the achievable communication rate of device $n$ in round $t$. 

\begin{algorithm}[t]
	\caption{Device Scheduling and Power Allocation Algorithm for TCS-H}\label{AlgoSchedule}
	\begin{algorithmic}[1]
		\State \textbf{Input:} $\bar{P}_{t,n}$, $P_{t,n}^{\text{[g]}}$, $h_{t,n,m}$, $\mathcal{N}_t=\emptyset$, $\mathcal{M}_t=\emptyset$.
		\State \textbf{Device scheduling:}
		\For {$n\in\mathcal{N}$}
		\State If $P_{t,n}^{\text{[g]}}\leq \alpha_{t,n}\bar{P}_{t,n}U_{t,n}^{\text{[g]}}$, then $\mathcal{N}_t=\mathcal{N}_t\cup \{n\}$.
		\EndFor
		\State \textbf{Sub-channel assignment and power allocation:}
		\State Initial sub-channel assignment by solving a bottleneck matching problem. 
		\State For $\forall n\in\mathcal{N}_t$, calculate rate $R_{t,n}$ under the current sub-channel assignment.
		\State Update $\mathcal{M}_t=\{m|\beta_{t,n,m}=1, m\in\mathcal{M}\}$. Let $\mathcal{M}'_t=\mathcal{M}-\mathcal{M}_t$ and $\mathcal{N}'_t=\mathcal{N}_t$.
		\While {$\mathcal{M}'_t\neq\emptyset$ and $\mathcal{N}'_t\neq\emptyset$}
		\State $n^{\dagger}=\arg\min_{ n\in\mathcal{N}'_t} R_{t,n}$.
		\State $m^{\dagger}=\arg\max_{ m\in\mathcal{M}'_t} r_{t,n^{\dagger},m}$.
		\State Calculate potential gain $\Delta R_{t,n^{\dagger}}$ according to \eqref{delta_r}.
		\If {$\Delta R_{t,n^{\dagger}}> 0$}
		\State $\beta_{t,n^{\dagger},m^{\dagger}}=1$, $\mathcal{M}_t=\mathcal{M}_t\cup \{m^{\dagger}\}$, $\mathcal{M}'_t=\mathcal{M}-\mathcal{M}_t$. Update $R_{t,n^{\dagger}}$.
		\Else 
		\State $\mathcal{N}'_t=\mathcal{N}'_t-\{n^{\dagger}\}$.
		\EndIf
		\EndWhile 
		\State For $\forall n\in\mathcal{N}_t$, use water-filling algorithm for the rate-optimal power allocation, as shown in \eqref{water_filling}.
		\State \textbf{Output:} $\mathcal{N}_t$, $\boldsymbol{\beta}_t$, $\boldsymbol{P}_t$, $U_t$
	\end{algorithmic}
\end{algorithm}


Problem $\mathcal{P}1$ is a mixed-integer non-linear programming problem, which is difficult to solve. Inspired by the channel and power allocation policies for orthogonal frequency division multiplexing (OFDM) systems \cite{gao2008efficient,Huang2009joint}, we decouple sub-channel assignment from power allocation, and solve $\mathcal{P}1$ in the following three steps. Note that the number of sub-channels is in general larger than that of the devices, and thus we assume $M\geq N$ in the following.

\emph{1) Initial Sub-Channel Assignment:}
The BS first assigns one sub-channel $m\in\mathcal{M}$ to each device $n\in\mathcal{N}_t$, such that the minimum achievable rate of devices is maximized. Let $r_{t,n,m}=\log_2\left( 1+\frac{\bar{P}_{t,n}|h_{t,n,m}|^2}{\sigma_0^2}  \right)$ be the communication rate when a single sub-channel $m$ is allocated to device $n$. The sub-channels $\mathcal{M}$ and devices $\mathcal{N}_t$ form a complete undirected bipartite graph, where the weight of each edge between $m\in\mathcal{M}$ and $n\in\mathcal{N}_t$ is $r_{t,n,m}$. The assignment problem is called a \emph{bottleneck matching} or \emph{max-min matching} problem, whose optimal solution can be achieved by the threshold policy \cite{bipartite} with complexity $O(M^2N_t^2)$. The key idea of the threshold policy is to activate edges whose weights are larger than a progressively decreasing threshold, and find a maximum cardinality matching given the active edges.
The policy terminates when each device is assigned a sub-channel.

\emph{2) Remaining Sub-Channel Assignment:} 
The remaining $M-N_t$ sub-channels, denoted by $\mathcal{M}'_t$, are further assigned in a heuristic manner. 
Let $\mathcal{N}'_t$ be the set of devices whose sum rates $R_{t,n}$ can be improved if more sub-channels are assigned, with $\mathcal{N}'_t= \mathcal{N}_t$ in the beginning. As shown in Lines 10-19 in Algorithm \ref{AlgoSchedule}, in each iteration, we find device $n^{\dagger}\in\mathcal{N}'_t$ with the minimum sum rate, and the best remaining channel $m^{\dagger}\in\mathcal{M}'_t$ associated with $n^{\dagger}$. The potential gain of the sum rate is approximated under equal power allocation, which is given by 
\begin{align}\label{delta_r}
\Delta R_{t,n^{\dagger}}&=\sum_{m=1}^{M} \beta'_{t,n^{\dagger},m}\log_2\left( 1+\frac{\bar{P}_{t,n^{\dagger}}|h_{t,n^{\dagger},m}|^2}{\sigma_0^2\sum_{m=1}^{M} \beta'_{t,n^{\dagger},m}}\right) \nonumber\\
&-\sum_{m=1}^{M} \beta_{t,n^{\dagger},m}\log_2\left( 1+\frac{\bar{P}_{t,n^{\dagger}}|h_{t,n^{\dagger},m}|^2}{\sigma_0^2\sum_{m=1}^{M} \beta_{t,n^{\dagger},m}}\right), 
\end{align}
where $\beta'_{t,n^{\dagger}, m^{\dagger}} =1$, and $\beta'_{t,n^{\dagger},m}=\beta_{t,n^{\dagger},m}$ for $\forall m\neq m^{\dagger}$.  

If $\Delta R_{t,n^{\dagger}}$ is positive, then we assign sub-channel $m^{\dagger}$ to device $n^{\dagger}$. Otherwise, assigning more sub-channels is not beneficial to rate increment for device $n^{\dagger}$, which is thus removed from $\mathcal{N}'_t$.
The process is terminated when all the sub-channels are assigned, or no device can benefit from having more sub-channels.

\emph{3) Power Allocation:}
Given sub-channel assignment $\boldsymbol{\beta}_t$, we finally implement the water-filling algorithm for each device to optimize power allocation for rate maximization. For $\forall n\in\mathcal{N}_t$, power is allocated according to
\begin{align} \label{water_filling}
P_{t,n,m}^{\text{[l]}}=
\begin{cases}
\left(\frac{1}{\lambda_{t,n}}-\frac{\sigma_0^2}{|h_{t,n,m}|^2}\right)^+, &\text{if} ~\beta_{t,n,m}=1, \\
0, &\text{otherwise},
\end{cases}
\end{align}
where $\lambda_{t,n}$ is chosen such that $\sum_{m=1}^{M} P_{t,n,m}^{\text{[l]}} =\bar{P}_{t,n}$,
$(x)^+=x$ if $x>0$, and $(x)^+=0$ otherwise. 

Finally, Algorithm \ref{AlgoSchedule} outputs the device scheduling $\mathcal{N}_t$, sub-channel assignment $\boldsymbol{\beta}_t$, power allocation $\boldsymbol{P}_t$, and the total number of communication slots $U_t$ consumed in round $t$. 



\section{Convergence Analysis}
In this section, we provide a convergence guarantee for the proposed TCS-H algorithm. 


\begin{lemma} \label{lemma-qsparse}
	For $\forall \x\in\mathbb{R}^d$, using a global mask $\m^{\rm{[g]}}$ and a local mask $\m^{\rm{[l]}}$ for sparsification, and applying a $q$-bit stochastic quantization as \cite{QSGD} to the local sparsified vector,  
	\begin{align}
		&\mathbb{E}\left[\lVert\x-\x\circ \m^{\rm{[g]}} - Q(\x\circ \m^{\rm{[l]}},q) \rVert_2^2\right]  \nonumber\\
		&~~~~~~~~~~~~~~
		\leq \left[1-\left( 1- \frac{K^{\rm{[l]}}}{2^{2q-2}} \right)\frac{K^{\rm{[g]}}+K^{\rm{[l]}}}{d}\right]\lVert\x\rVert_2^2,
	\end{align}
	where $K^{\rm{[g]}}$ and $K^{\rm{[l]}}$ are the numbers of non-zero elements in $\m^{\rm{[g]}}$ and $\m^{\rm{[l]}}$, respectively, $K^{\text{[l]}}<2^{2q-2}$, and the expectation is taken over sparsification and stochastic quantization.
\end{lemma}

To facilitate the convergence analysis, we follow the literature \cite{Basu2019Qsparse,QSGD,Stich2018Sparsified}, and assume that the local loss function $F_n(\w)$ is $L$-smooth and $\mu$-strongly convex, i.e., for $\forall \v,\w \in \mathbb{R}^d$ and $n\in\mathcal{N}$, 
$F_n(\v)-F_n(\w)\leq \langle\nabla F_n(\w), \v-\w\rangle+\frac{L}{2}\left\lVert \v-\w\right\rVert_2^2$,
$F_n(\v)-F_n(\w)\geq \langle\nabla F_n(\w), \v-\w\rangle+\frac{\mu}{2}\left\lVert \v-\w\right\rVert_2^2$. Here, $\langle \x, \y\rangle$ denotes the inner product of $\x$ and $\y$.
We also assume that the local stochastic gradient has bounded $l_2$-norm, i.e., 
$\mathbb{E}_{\x_n\sim\mathcal{D}_n}\left[\left\lVert \nabla F_n\left(\w_{t,n,i};\x_n\right)\right\rVert_2^2\right]\leq G^2$, $\forall n,t,i$.

Define $\gamma\triangleq\left( 1- \frac{K^{\rm{[l]}}}{2^{2q-2}} \right)\frac{K^{\rm{[g]}}+K^{\rm{[l]}}}{d}$, $A_1 \geq\frac{4a\gamma(1-\gamma^2)}{a\gamma-4H}$, where $a>0$, and the following auxiliary variables:
$V_1=\left(\frac{7\mu H}{2}+6HL\right)\frac{\sigma_0^2K^{[\text{g}]}}{N^2}$, $V_2=\frac{HG^2}{BN}$, 
$V_3=\left(\frac{3\mu}{2}+\frac{9L}{2}\right)G^2(H-1)H +(14\mu+24L)\frac{A_1H^3G^2}{\gamma}$.


The convergence of the proposed TCS-H algorithm under full device participation is given as follows, which mainly shows the impact of the channel noise and model compression.

\begin{theorem} \label{convergence-TCS}
	Given the initial global model $\w_0$, let $\sigma_t={(a+t)^2}$, the learning rate $\eta_t=\frac{8}{\mu H(a+t)}$, and define $\omega_t=(a+t)^2$, when $a>\max\left\{\frac{4H}{\gamma}, \frac{32L}{\mu} \right\}$,
	\begin{align} \label{convergence-TCS-eq}
	F(\bar{\w}_T)&-F^* \leq ~~\frac{L a^3}{4S_T}\lVert \w_0-\w^* \rVert_2^2 + \frac{LT^2}{\mu H S_T}V_1 \nonumber\\
	&~~~+\frac{8LT(T+2a+1)}{\mu^2 H^2S_T }V_2 +\frac{64LT}{\mu^3 H^3 S_T}V_3,
	\end{align}
	where $S_T=\sum_{t=1}^{T} \omega_t$, $\bar{\w}_T=\frac{1}{S_T}\sum_{t=1}^{T}\omega_t\w_{t-1}$, $\w^*$ and $F^*$ are the optimal global model and minimum global loss, respectively.
\end{theorem}
The proof uses Lemma \ref{lemma-qsparse} and follows the perturbed iterate framework in \cite{Basu2019Qsparse,Stich2018Sparsified}. We remark two major differences compared with the proof in \cite{Basu2019Qsparse}.
One is that we further take into account the wireless channel noise, which introduces a new term $\frac{LT^2}{\mu H S_T}V_1$ in the bound. The other is that we consider the convergence of the global model, by folding up the local iterations in the proof.
As $S_T$ is at the scale of $O(T^3)$, the bound given in \eqref{convergence-TCS-eq} indicates the convergence of TCS-H.

\section{Experiments}

In this section, we evaluate the proposed TCS-H algorithm for FEEL by considering an image classification task on the CIFAR-10\footnote{\url{https://www.cs.toronto.edu/~kriz/cifar.html}} dataset. There are $N=20$ devices and $M=25$ sub-channels. We consider both independent and identically distributed (i.i.d.) and non-i.i.d. local data across devices.
For the i.i.d. case, the whole training dataset is partitioned into $N$ disjoint subsets uniformly at random. For the non-i.i.d. case, each device stores $2$ out of $10$ classes of data samples. 

A convolutional neural network (CNN) with the same structure as \cite{Sun2021JSAC} is trained.
The total number of parameters of the considered CNN is 258898.
The mini-batch size is 64, and the learning rate is 0.05. Regarding the wireless channel, we consider Rayleigh fading with scale parameter 1, and additive white Gaussian noise with variance $\sigma_0^2=10^{-6}$. The average power constraint is $\bar{P}_n=5\mathrm{mW}$, $\forall n$, and the power scalar is $\sigma_t=5$, $\forall t$.

We compare the proposed TCS-H algorithm with two benchmarks.
1) Top-$K$: each device sparsifies the local model difference $\g_{t,n}^{\text{[ec]}}$ by keeping $K=K^{\text{[g]}}+K^{\text{[l]}}$ elements with the largest absolute values. Orthogonal digital communication is used for global aggregation.
2) TCS-D \cite{Emre2021spars}: it uses TCS with the same global and local sparsities $K^{\text{[g]}}$ and $K^{\text{[l]}}$, but both $\widetilde{\g}_{t,n}^{\text{[g]}}$ and $\widetilde{\g}_{t,n}^{\text{[l]}}$ are transmitted via orthogonal digital communications. 
For both Top-$K$ and TCS-D, the number of bits to transmit each element of $\widetilde{\g}_{t,n}^{\text{[g]}}$ and $\widetilde{\g}_{t,n}^{\text{[l]}}$ is the same as that of $\widetilde{\g}_{t,n}^{\text{[l]}}$ in TCS-H, represented by $q$.

\begin{figure}[!t]
	\centering
	\includegraphics[width=0.5\textwidth]{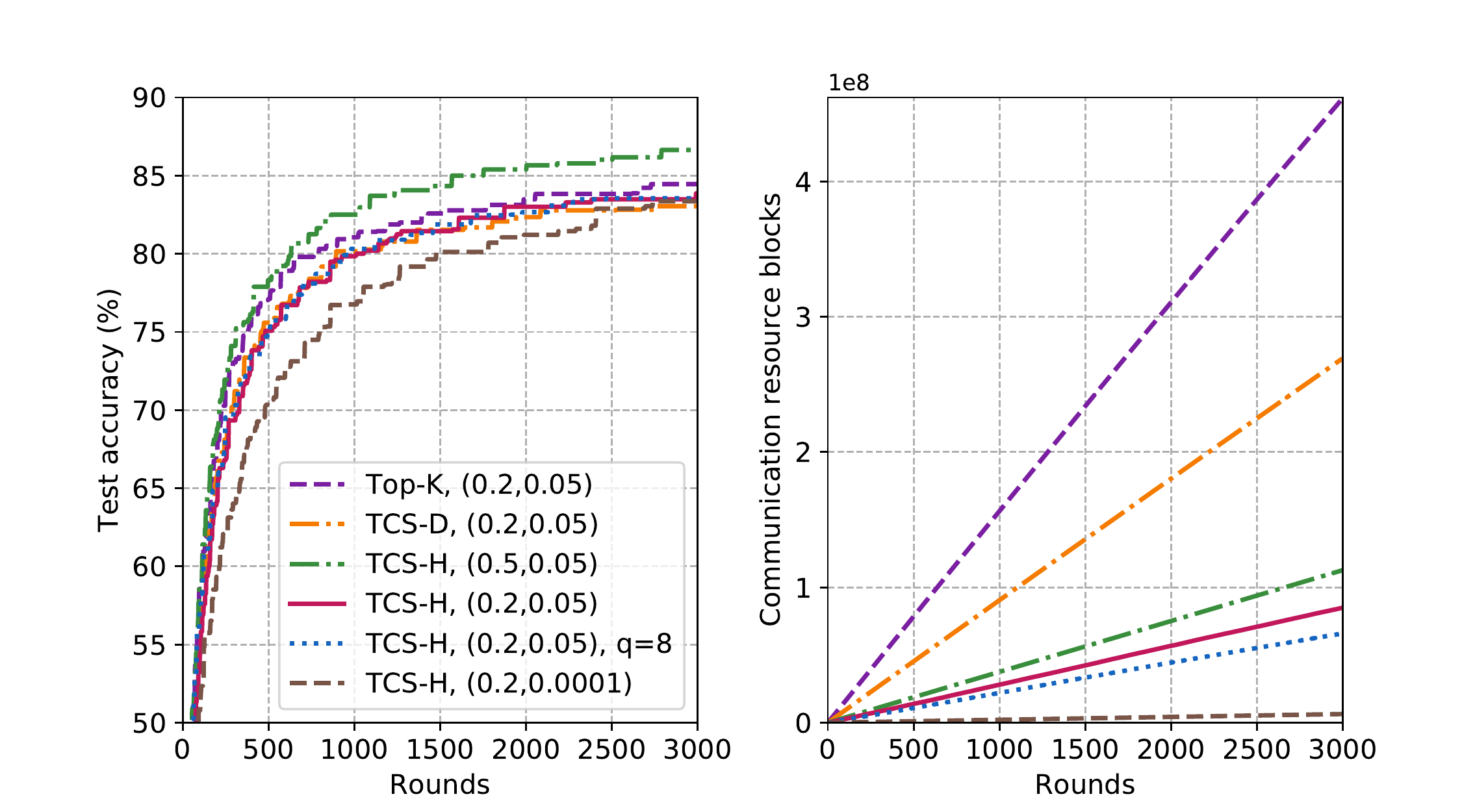}
	\vspace{-5mm}	
	\caption{Model accuracy (left) and the required communication resources (right) of different algorithms  on CIFAR-10 dataset with i.i.d. data.  }
	\label{acc-iid}
\end{figure}

Fig. \ref{acc-iid} shows the model accuracy achieved by different algorithms in the i.i.d. case, and the corresponding number of communication resource blocks $\sum_{\tau=1}^{t}U_{\tau}M$ consumed. We run each algorithm for $3000$ rounds, with $H=10$ local iterations in each round. 
In the legend, $(\phi^{\text{[g]}}, \phi^{\text{[l]}})$ shows the global and local sparsities, i.e., $\phi^{\text{[g]}} =\frac{K^{\text{[g]}}}{d}$ and $\phi^{\text{[l]}}=\frac{K^{\text{[l]}}}{d}$. The default value of $q$ is 16, unless specified. 
Given $\phi^{\text{[g]}}=0.2$ and $\phi^{\text{[l]}}=0.05$, the average fraction of devices scheduled by TCS-H is 0.65, and thus for Top-$K$ and TCS-D, we randomly schedule $13$ devices in each round.

When $\phi^{\text{[g]}}=0.2$ and $\phi^{\text{[l]}}=0.05$, we find that TCS can maintain similar accuracy as Top-$K$, while greatly reduce the communication cost without transmitting the global masks. Meanwhile, the proposed TCS-H algorithm with OAC can further save $68\%$ communication resources compared with the TCS-D algorithm with pure orthogonal aggregation. 
Comparing TCS-H with different sparsities, we find that even when $\phi^{\text{[l]}}=0.0001$, the TCS-H algorithm can still achieve a similar accuracy as that with $\phi^{\text{[l]}}=0.05$. This validates the strong temporal correlation of the model, and thus the feasibility of the TCS-H algorithm.
Last but not least, TCS-H can outperform the Top-$K$ algorithm in model accuracy and communication efficiency simultaneously. For example, when $\phi^{\text{[l]}}=0.05$, the model accuracy achieved by TCS-H with $\phi^{\text{[g]}}=0.5$ is $2.18\%$ higher than that of Top-$K$ with $\phi^{\text{[g]}}=0.2$, while the number of communication resource blocks consumed by TCS-H is $75\%$ fewer.

\begin{figure}[!t]
	\centering
	\includegraphics[width=0.5\textwidth]{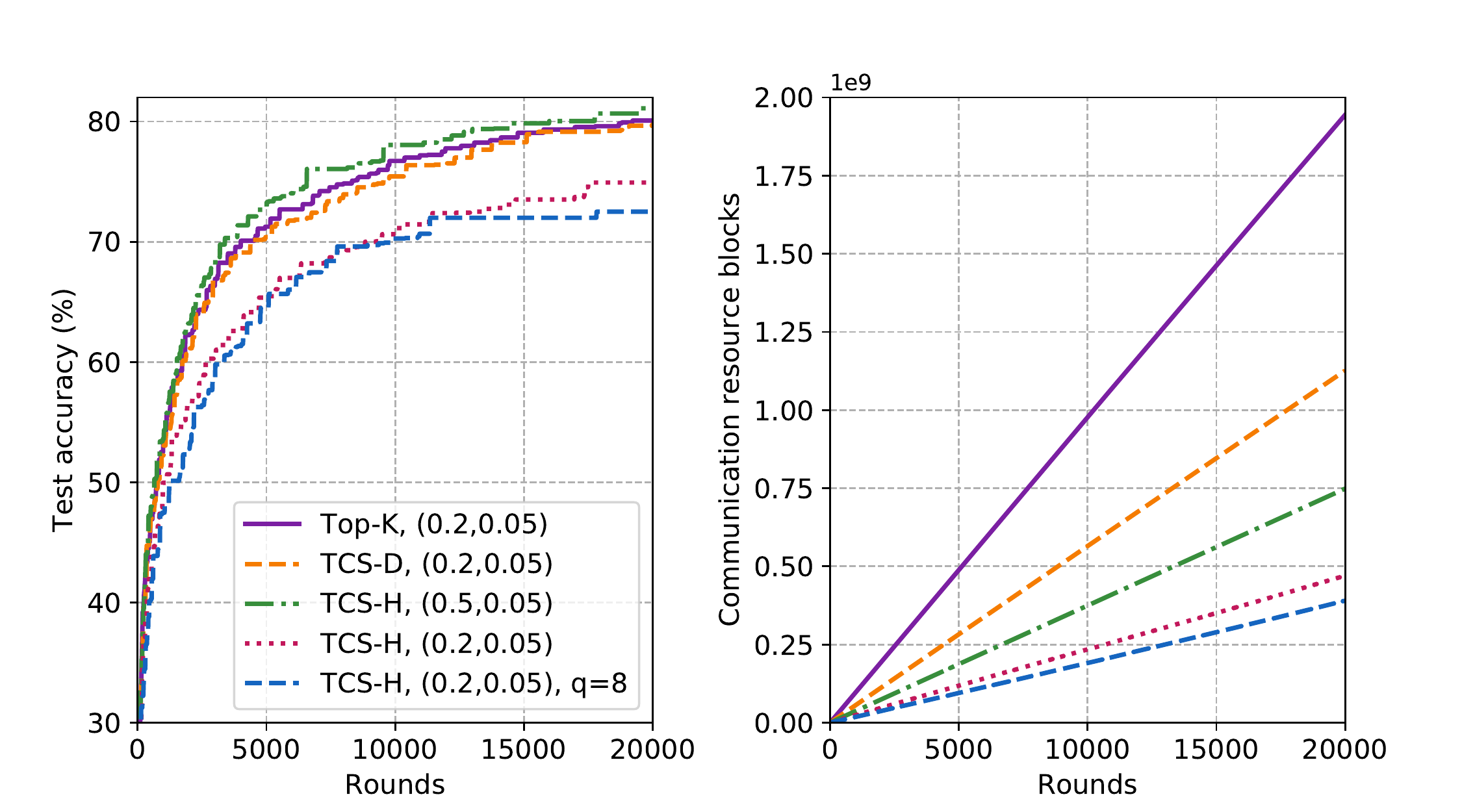}
	\vspace{-5mm}	
	\caption{Model accuracy (left) and the required communication resources (right) of different algorithms on CIFAR-10 dataset with non-i.i.d. data.  }
	\label{acc-noniid}
	\vspace{-2mm}
\end{figure}

In the non-i.i.d. case, we run each algorithm for $20000$ rounds, with one local iteration in each round.
$11$ devices are scheduled by TCS-H on average, which is also set for the Top-$K$ and TCS-D benchmarks.
As shown in Fig. \ref{acc-noniid}, under non-i.i.d. data, the proposed TCS-H algorithm can still achieve a higher model accuracy than the Top-$K$ and TCS-D benchmarks with fewer communication resources, which is similar to the i.i.d. case. Moreover, TCS-H has a significant gain when the total communication resource is limited. For example, if the number of communication resource blocks is limited to $7.5\times10^8$, then TCS-H, TCS-D and Top-$K$ can train the CNN model for 20000, 13290 and 7664 rounds, respectively. Accordingly, the accuracies achieved by the three algorithms are $81.1\%$, $77.7\%$ and $74.5\%$. TCS-H increases the model accuracy by $6.6\%$ compared with Top-$K$ sparsification. Such result validates the superior communication efficiency of TCS-H.

\section{Conclusions}
In this work, we have proposed a TCS-H algorithm to improve the communication efficiency of FEEL. By exploiting the temporal correlations among model parameters, a global mask is constructed in each training round, which enables the model aggregation over-the-air. The important parameters of each individual device are further explored and aggregated via orthogonal digital communication. 
Experiments on CIFAR-10 dataset have shown that, the proposed TCS-H is extremely promising when the wireless communication resource is limited, with a $6.6\%$ accuracy improvement compared to the Top-K sparsification via orthogonal model aggregation.
Future directions include the design of online device scheduling policies, and the optimization of the sparsity and quantization levels under various kinds of resource constraints.

%

\end{document}